\documentclass[twocolumn,twoside,slac_two]{revtex4}
\usepackage{graphicx}
\usepackage{fancyhdr}
\begin{document}

\title{Modeling the high-energy radiation in gamma-ray binaries}

%

\author{B. Cerutti, G. Dubus, G. Henri}
\affiliation{Laboratoire d'Astrophysique de Grenoble, UMR 5571 CNRS, Universit\'e
Joseph Fourier, BP 53, 38041 Grenoble, France}

\begin{abstract}

Gamma-ray binaries are orbital modulated gamma-ray sources in the Galaxy detected both at GeV and TeV energies. The high-energy radiation may come from the interaction of energetic electrons injected by a young pulsar and photons from the massive companion star. We present a model for the production of high-energy radiation where emission, absorption and pair cascading are considered. New observations of LS~I~$+61^{\rm{o}}303$ by the {\em Fermi Space Telescope} revealed an exponential cut-off in the spectrum at a few GeV, inconsistent with gamma-gamma absorption. Electrons radiating at GeV and TeV have probably two different origins. We investigate whether the emission from the unshocked pulsar wind explains the GeV component in LS~I~$+61^{\rm{o}}303$.

\end{abstract}

\maketitle

\thispagestyle{fancy}


\section{Introduction}

Three binary systems have been discovered to produce non-thermal radiation up to very-high energies in the Galaxy (see also the contribution of J.~Holder in these proceedings) namely\footnote{Note that the detection of Cyg~X-3 by {\em Fermi} has been anounced at the Symposium (see S.~Corbel's proceeding). Note also that TeV emission from Cyg~X-1 has been reported by MAGIC \cite{2007ApJ...665L..51A}.}: LS~I~$+61^{\rm{o}}303$ \cite{2006Sci...312.1771A,2008ApJ...679.1427A}, LS~5039 \cite{2005Sci...309..746A} and PSR B1259$-$63 \cite{2005A&A...442....1A}. HESS~J0632$+$057, an unidentified gamma-ray source serendipitously discovered by HESS \cite{2007A&A...469L...1A}, may be the fourth system known \cite{2009ApJ...690L.101H}. These systems are composed of a B$e$ or O companion star and a compact object in an eccentric orbit. Except for PSR B1259$-$63 where a 48 ms radio pulsar has been detected \cite{1992ApJ...387L..37J}, the nature of the compact star remains unknown in the other two binaries.

These systems are known today as ``gamma-ray binaries'' as their non stellar luminosity peaks in gamma rays \cite{2006A&A...456..801D}. The main feature of these objects is probably the orbital modulation of the radiated gamma-ray flux. This property has been first uncovered at TeV energies by Cherenkov telescopes \cite{2006A&A...460..743A,2009ApJ...693..303A,2009A&A...507..389A}. The {\em Fermi Space Telescope} has recently reported the first detections of the orbital modulation at GeV energies \cite{2009ApJ...701L.123A,2009ApJ...706L..56A} (see Richard Dubois' proceeding in this volume for more details).

In this proceeding, we report on our modeling investigations of the high and very-high energy radiation produced in gamma-ray binaries in the framework of the pulsar wind nebula scenario. In this model, the non-thermal radiation is emitted by a cooling plasma of relativistic electron-positron pairs injected and accelerated by a young pulsar (see Section~2). We present in Section~3 our modeling of the GeV and TeV modulation in LS~5039 where emission, absorption and pair cascading are considered. In Section~4, we discuss the new observations of LS~I~$+61^{\rm{o}}303$ by {\em Fermi} and investigate whether the emission from an unshocked pulsar wind can explain the GeV component.

\section{The pulsar wind nebula scenario: the big picture}

\begin{figure*}[t]
\centering
\includegraphics[width=125mm]{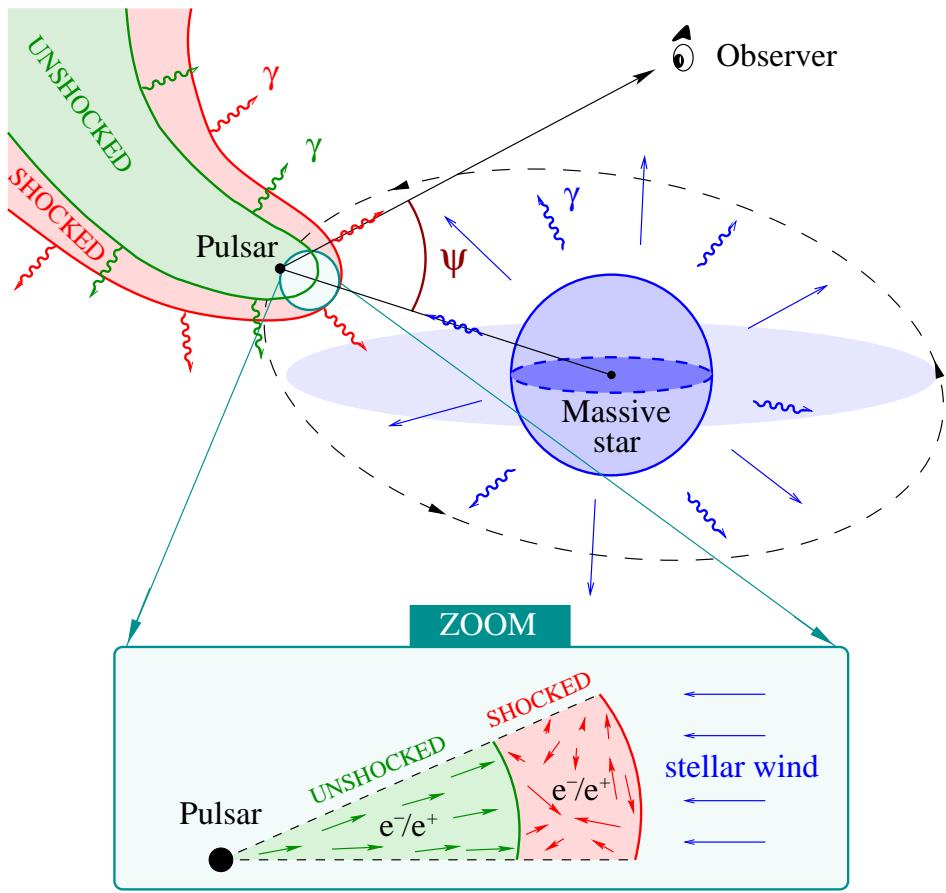}
\caption{This diagram depicts the pulsar wind nebula scenario in gamma-ray binaries. The massive star wind collides and confines the relativistic pulsar wind in a collimated outflow. Non-thermal radiation originates from the `unshocked' and `shocked' part of the pulsar wind. The observed gamma-ray flux depends on the position of both stars with respect to the observer (viewing angle $\psi$).}
\label{fig1}
\end{figure*}

Maraschi \& Treves \cite{1981MNRAS.194P...1M} suggested that the radio to gamma-ray radiation could be powered by a young ($\sim 10^4$~-~$10^5$ yr) fast-rotating pulsar in LS~I~$+61^{\rm{o}}303$ (scenario originally applied to Cyg~X-3 \cite{1977A&A....55..155B}). The same idea has been recently revisited and proposed to be at work in the other gamma-ray binaries \cite{2006A&A...456..801D}. In this scenario (see the sketch in Fig.~\ref{fig1}), the pulsar injects energetic electrons-positrons pairs in a relativistic wind. Particles propagate up to the termination shock where the pulsar wind momentum is balanced by the ram pressure of the massive star wind. The stellar wind can possibly confine the pulsar wind and a comet-like structure develops, spiraling around the system due to the orbital motion. At the shock, leptons are randomized, re-accelerated and produce non-thermal radiation.

In the `shocked' wind, pairs emit synchrotron radiation and up-scatter the soft photons from the massive star to high energies by inverse Compton scattering. In the `unshocked' wind region, pairs radiate only {\em via} inverse Compton scattering since upstream the termination shock, the magnetic field is frozen into the flow of particles. No synchrotron radiation is produced. The annihilation of gamma rays with the stellar photons can also be of major importance in some tight binaries such as LS~5039 (see next section). The production of gamma rays by inverse Compton scattering is not isotropic in gamma-ray binaries because the soft radiation field set by the massive star is anisotropic with respect to the location of electrons. Hence, the gamma-ray flux depends on the relative position of both stars and the observer (Fig.~\ref{fig1}).

\section{Modeling the high-energy radiation in LS~5039}

LS~5039 is composed of a O6.5V star in a compact 3.9~day orbit with its compact object. The orbital parameters are known with good accuracy (the latest solution can be found in \cite{2009ApJ...698..514A}). The absence of a B$e$ equatorial wind component makes this system ideal for modeling. TeV observations of this binary by HESS have shown a stable modulation of the gamma-ray flux with a maximum close to inferior conjunction and a minimum at superior conjunction \cite{2006A&A...460..743A}. The main behavior of this modulation can be accounted for by the effect of photon-photon annihilation $\gamma+\gamma\rightarrow e^{+}+e^{-}$ (see {\em e.g.} \cite{2006A&A...451....9D}). In the pulsar wind nebula scenario, the high-energy radiation has two distinct contributions.

The first component originates from the shocked wind where pairs are assumed to be isotropized and injected with a power-law energy distribution cooling down via synchrotron radiation and inverse Compton scattering. Combining emission and $\gamma\gamma$-absorption of gamma rays, HESS observations can be reasonably well explained. Some physical quantities at the shock can be tightly constrained such as the magnetic field $B=0.8\pm0.2$~G, the spectral index of the injected particle distribution $p=2\pm 0.3$ and the pulsar spin down power $L_P=10^{36}$ erg.$\rm{s^{-1}}$(see \cite{2008A&A...477..691D} for more details).

This simple model can reproduce most of the spectral and temporal features at high energies but underestimates the flux observed by HESS at orbital phases where the gamma-ray flux is highly absorbed, {\em i.e.} close to superior conjunction. The development of a cascade of pairs has been proposed to solve this discrepancy \cite{2006A&A...460..743A}. The large density of pairs produced by photon-photon absorption in the system can reprocess a significant amount of the absorbed power. If the ambient magnetic field is high enough, pairs are locally isotropized and radiate in all directions, the cascade is three-dimensional \cite{1997A&A...322..523B}. But the magnetic field should not exceed a few Gauss (5~G in LS~5039) otherwise particles will emit preferentially synchrotron radiation whose energy is too small to create new pairs in the system. Hence the cascade is quenched.

\begin{figure}
\centering
\resizebox{\hsize}{!}{\includegraphics{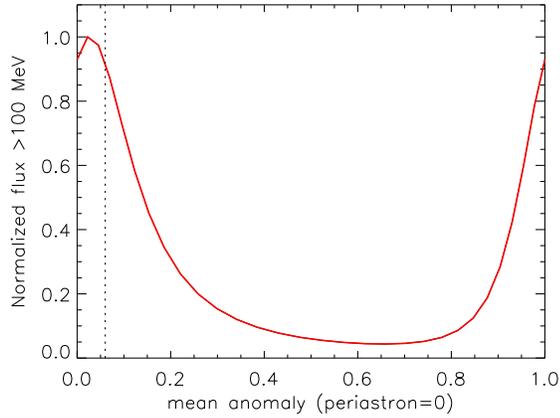}} 
  \caption{Computed lightcurve in the {\em Fermi} energy band (flux $>100$~MeV) for LS~5039. The dotted line indicates the orbital phase at superior conjunction. The inclination of the orbit is $i=40^{\rm{o}}$.}
\label{fig2}
\end{figure}

\begin{figure}
\centering
\resizebox{\hsize}{!}{\includegraphics{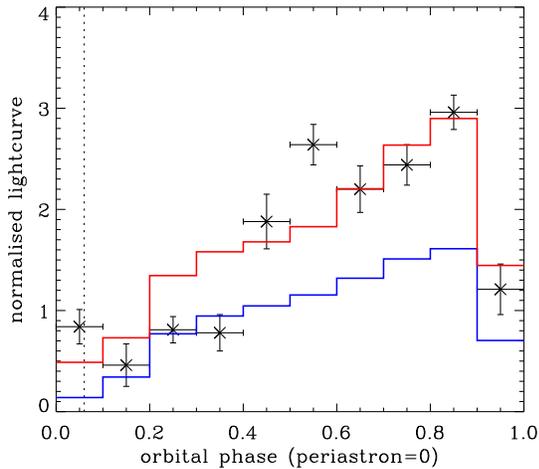}} 
  \caption{Computed lightcurve in the HESS energy band (flux $>1$~TeV) for LS~5039. The red line gives the total flux with the cascade contribution. The blue line shows the absorbed primary flux for comparison. Lightcurves are averaged over an orbital phase bin $\Delta\phi=0.1$ and are compared with HESS data points taken from \cite{2006A&A...460..743A}. The dotted line indicates the orbital phase at superior conjunction. The inclination of the orbit is $i=40^{\rm{o}}$.}
\label{fig3}
\end{figure}

We present in Figs.~\ref{fig2}-\ref{fig3} the computed lightcurves in the {\em Fermi} and HESS energy bands where the full calculation of a three-dimensional pair cascade radiation is taken into account using a Monte Carlo code (Cerutti et al., in preparation). The cascade contribution is significant at every orbital phase and dominates close to superior conjunction, giving a consistent flux with HESS observations (Fig.~\ref{fig3}). The anti-correlation between the GeV and the TeV lightcurves is due to pair production (see \cite{2008A&A...477..691D} for more details).

The second high-energy radiation component originates from the unshocked pulsar wind. The inverse Compton scattering on a mono-energetic radial wind of relativistic pairs produces a sharp line-like component centered at an energy set by the injected Lorentz factor of the wind (see \cite{2008A&A...488...37C} for more details). In LS~5039, this contribution should have a strong spectral signature but is not observed at high energy suggesting that the Lorentz factor of the wind might be low ($<10^4$) or very high ($>10^7$). Alternatively, the pulsar wind might have a `striped' structure and remain highly magnetized up to the termination shock (see for instance \cite{2007astro.ph..3116K}). Another scenario has been proposed where the very-high energy radiation is dominated by the unshocked pulsar wind \cite{2008APh....30..239S}.

\section{New modeling challenges: LS~I~+61$^{\rm{o}}$303 seen by {\em Fermi}}

The {\em Fermi} collaboration has recently reported the detection of the gamma-ray binary LS~I~$+61^{\rm{o}}$303 and the first detection of an orbital modulation at GeV \cite{2009ApJ...701L.123A}. The lightcurve peaks close to periastron and is minimum close to apastron. The measured spectrum is a power-law with a 2.2 photon index and an exponential cut-off at 6.3 GeV. This energy cut-off is too low to be due to pair production on stellar photons. Electrons radiating at GeV and TeV have probably two different origins. It is worthwhile to note that this spectrum is very similar to those observed in gamma-ray pulsars. {\em Fermi} might be observing the magnetospheric emission from the pulsar in the binary system. In this case, the origin of the orbital modulation remains unclear.

\begin{figure}
\centering
\resizebox{\hsize}{!}{\includegraphics*{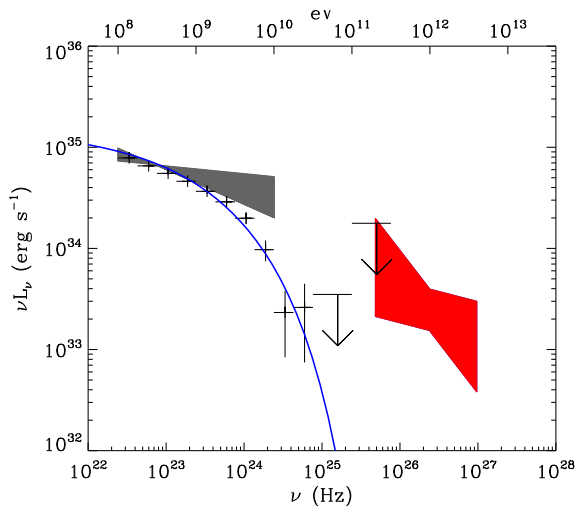}} 
  \caption{Orbit-averaged spectral signature of an unshocked pulsar wind in LS~I~$+61^{\rm{o}}$303 (blue line). Observations are overplotted for comparison namely EGRET (grey bowtie, \cite{1999ApJS..123...79H}), {\em Fermi} (black crosses, \cite{2009ApJ...701L.123A}) and MAGIC (red bowtie, \cite{2006Sci...312.1771A}).}
\label{fig4}
\end{figure}

\begin{figure}
\centering
\resizebox{\hsize}{!}{\includegraphics{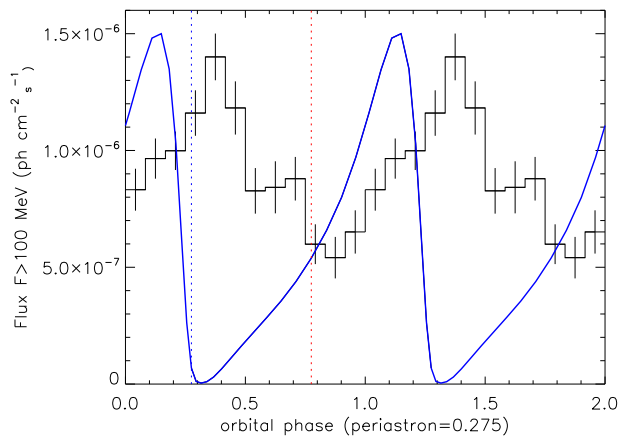}} 
  \caption{Theoretical lightcurve (flux $>100~$MeV) computed in LS~I~$+61^{\rm{o}}$303 (blue solid line). The gamma-ray flux is assumed to originate from the unshocked pulsar wind. The dotted lines show the orbital phase at periastron (blue) and apastron (red). Orbital parameters are taken from \cite{2009ApJ...698..514A}. The folded lightcurve measured by {\em Fermi} \cite{2009ApJ...701L.123A} is overplotted for comparison (black solid line).}
\label{fig5}
\end{figure}

Alternatively, the high-energy radiation component might come from the unshocked pulsar wind (for another alternative solution, see the contribution by M. Chernyakova in these proceedings). We investigate here whether this possibility could provide a viable explanation for the {\em Fermi} observations. Figure~\ref{fig4} shows the computed inverse Compton radiation produced by the interaction of the stellar photons with the pairs in the unshocked wind. Pairs are injected with a soft power-law distribution with an index $\approx 3.1$ and an exponential cut-off at $\approx 25$~GeV. To account for the observed flux, the pulsar luminosity should be high $L_p>10^{37}$ erg.$\rm{s^{-1}}$.

Figure~\ref{fig5} gives the expected GeV modulation using the orbital parameters found in \cite{2009ApJ...698..514A}. The lightcurve peaks just before periastron, at an orbital phase ($\phi\approx 0.15$) where both the soft photon density and the angle of interaction between the pairs and the photons is high (close to head-on collision). The peak is followed by a steep decline of the gamma-ray flux. The lightcurve is minimum just after periastron ($\phi\approx 0.3$) since the interaction between the pairs and the stellar photons occurs almost rear-end, hence highly inefficient even though the soft photon density remains high at this phase. Nonetheless, this model cannot account for the observed GeV modulation. A significant shift in phase ($\Delta\phi\approx+0.25$) of the theoretical lightcurve would be needed to match observations, but there is no obvious physical motivation for this yet.

\section{Conclusion}

Leptonic model can explain the general features observed at high and very-high energy in gamma-ray binaries. In LS~5039, the GeV and TeV modulation are well reproduced but the interplay between emission, absorption and pair cascading is complex. {\em Fermi} measurements of the spectrum and the modulation in LS~I~$+61^{\rm{o}}$303 challenge simple models. Particles radiating at GeV and TeV energies may have two different origins. The GeV component might be the signature of an unshocked pulsar wind in the system. More investigations are required to understand the origin of the GeV modulation in LS~I~$+61^{\rm{o}}$303.

\bigskip 
\begin{acknowledgments}
This work is supported by the {\em European Community} contract ERC-StG-200911.
\end{acknowledgments}

\bigskip 





\end{document}